\documentclass{INTERSPEECH2023}

\interspeechcameraready 
\usepackage{multirow}
\usepackage{graphicx}  
\usepackage{subcaption} 
\usepackage[T1]{fontenc}

\title{{MuLanTTS: The Microsoft Speech Synthesis System for \\ Blizzard Challenge 2023}}
\name{Zhihang Xu, Shaofei Zhang, Xi Wang, Jiajun Zhang, Wenning Wei, Lei He, Sheng Zhao}
\address{
  Microsoft Azure Speech, China
}
\email{\{zhihangxu, shazh, xwang, jiajzhan, wennwei, helei, szhao\}@microsoft.com}

\begin{document}

\maketitle
 
\begin{abstract}

In this paper, we present MuLanTTS, the Microsoft end-to-end neural text-to-speech (TTS) system designed for the Blizzard Challenge 2023. About 50 hours of audiobook corpus for French TTS as hub task and another 2 hours of speaker adaptation as spoke task are released to build synthesized voices for different test purposes including sentences, paragraphs, homographs, lists, etc. Building upon DelightfulTTS, we adopt contextual and emotion encoders to adapt the audiobook data to enrich beyond sentences for long-form prosody and dialogue expressiveness. Regarding the recording quality, we also apply denoise algorithms and long audio processing for both corpora. For the hub task, only the 50-hour single speaker data is used for building the TTS system, while for the spoke task, a multi-speaker source model is used for target speaker fine-tuning. MuLanTTS achieves mean scores of quality assessment 4.3 and 4.5 in the respective tasks, statistically comparable with natural speech while keeping good similarity according to similarity assessment. The excellent quality and similarity in this year's new and dense statistical evaluation show the effectiveness of our proposed system in both tasks.

\end{abstract}

\section{Introduction}  
  
The Blizzard Challenge aims to advance the technologies in TTS~\cite{shen2018natural, ren2019fastspeech, ren2020fastspeech} by comparing and understanding different approaches. It has been organized annually since 2005~\cite{black2005blizzard}, with the basic task being to build high-quality TTS systems based on the speech database provided by the organizers. Participants use their developed systems to synthesize audio from the given test set, and the generated audio samples are used to evaluate the performance of different systems through subjective listening tests and objective metrics.  

End-to-end neural TTS has achieved significant naturalness quality improvements. 
Tacotron~\cite{shen2018natural,wang2017tacotron} and TransformerTTS~\cite{li2019neural} introduce an attention-based encoder-decoder acoustic model to predict mel-spectrograms from a character sequence, and a WaveNet model~\cite{oord2016wavenet} to synthesize waveforms from mel-spectrograms. These autoregressive models suffer from robustness issues and slow inference speed. RobuTrans~\cite{li2020robutrans} attempted to solve stability problems like repeated phones by using linguistic features and phone duration with an autoregressive Transformer. However, the inference is still slow due to the autoregressive nature of the model. FastSpeech~\cite{ren2019fastspeech} shifted towards non-autoregressive TTS by using a feed-forward transformer in the encoder and decoder for parallel generation and a duration predictor for phoneme duration. With the introduction of the Conformer~\cite{gulati2020conformer} architecture and rich variance adapter and prosody style~\cite{liu2021delightfultts,yi2022prosodyspeech} into TTS systems, it could achieve human-like quality in a specific speech corpus.


Compared with cascaded acoustic and vocoder models, single-stage end-to-end TTS models as new paradigms have been proposed recently. These include neural codec-based methods such as DelightfulTTS2~\cite{liu2021delightfultts}, VALLE~\cite{wang2023neural}, and Tortoise-TTS~\cite{betker2023better}, latent variable-based methods like VITS~\cite{kim2021vits} and NaturalSpeech~\cite{tan2022naturalspeech}, and pre-trained vector-based methods such as VQTTS~\cite{du2022vqtts}, SpearTTS~\cite{kharitonov2023speak}, and UniCATS~\cite{du2023unicats}. These models aim to reduce the training and inference mismatch and capture more detailed intermediate speech representations other than human-designed features like mel-spectrogram. With appropriate adversarial criteria and training strategy, the single-stage pure end-to-end model has room to achieve even higher fidelity.

With the fast-paced advancements in TTS, complex applications like paragraph TTS for long-form text synthesis with expressiveness and contextual variations have gained attention. Unlike single-sentence synthesis, it demands diverse prosody and smooth transitions. Contextual TTS addresses traditional TTS limitations by using extra contextual information for a more precise and expressive portrayal of the intended message. Recent progress include attention mechanisms~\cite{xiao2022improving,dai2019transformer}, conditional style prediction~\cite{lei2023context,xue2023m}, and the integration of pre-trained NLP models into TTS systems~\cite{xiao2023contextspeech, xue2023m,lei2023context}. 
According to this year's Blizzard Challenge with two tasks on 50 hours of audiobook synthesis and 2 hours of spoke speech adaptation, we design the model from two aspects: 1) incorporating rich text-level contextual and emotion encoders within additional text corpus, to capture the subtle nuances and variations that are characteristic of human speech in context variation. 2) building a multi-speaker source model to adapt to the spoke speaker. In addition, we have incorporated advanced speech enhancement algorithms to bolster the clarity and quality of the synthesized speech. These sophisticated algorithms are designed to minimize noise and other audio disruptions, which would benefit the perception of naturalness while noise may be amplified during training. Furthermore, our data preparation process has played a significant role in the transcription segmentation of long audios like audiobook data. 



Our system demonstrates superior performance in both tasks. In the MOS test, the mean scores are 4.3 for FH1 and 4.5 for FS1, both of which outperformed other TTS systems and are close to or matched the recording scores. In the Similarity MOS test, though FH1 task is quite hard due to the large data variation, our system still achieves the relatively highest mean score of 3.4 statistically comparing recordings. Meanwhile, in FS1, our System F obtains a mean SMOS score of 4.3 which surpasses the recording's score a little. This demonstrates the performance of our system in enhancing perceived quality and naturalness. For audio samples, please visit our website\footnote{Audio samples: \url{https://cognitivespeech.github.io/mulantts}}.

\section{System Structure}

\subsection{System Overview}
\begin{figure}[t]
  \centering
  \includegraphics[width=0.95\linewidth]{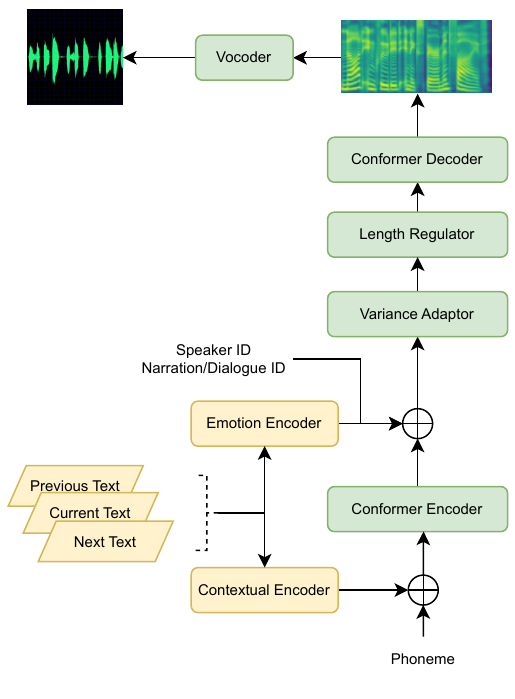}
  \caption{Our MuLanTTS system converts phonemes into a 16k mel-spectrogram using a Conformal-based model, extracts long and emotional text details with BERT-based encoders, and then applies HiFi-GAN to transform the mel-spectrogram into a 24kHz waveform.}
  \label{fig:overview}
\end{figure}

As shown in Figure \ref{fig:overview}, Our system employs the conventional TTS strategy, which comprises a frontend, an acoustic model, and a vocoder. The input text is firstly normalized and converted into phone sequences using the TTS frontend, and the resulting phone sequences are then fed into the acoustic model to generate a 16k mel-spectrogram. To improve the pronunciation of poly-phone and liaison phenomena in French, we utilize a BERT~\cite{devlin2018BERT} model to make predictions in a multitask fashion~\cite{pan2020unified}. Additionally, to model long context, text emotion, and narration-or-dialogue variance, we utilize BERT-based contextual encoder~\cite{xiao2023contextspeech} and emotion encoder to enhance the acoustic model's ability to generate more expressive speech\footnote{All pretrained BERTs used in liaison, polyphone, and contextual modeling are from https://huggingface.co/dbmdz/bert-base-french-europeana-cased}. Finally, the predicted mel-spectrogram is fed into a high-fidelity HiFi-GAN~\cite{kong2020hifi} to generate a 24kHz waveform.

\subsection{Liaison and Polyphone}
\subsubsection{Liaison}
Liaison is a French pronunciation rule that occurs between a word that ends in a consonant and one that starts with a vowel. It’s the practice of taking the last letter of one word and linking it to the beginning of the next word and primarily focuses on the final phonemes of 'n,' 't,' and 'z'. It usually is a major issue for French phonetizers.  Pontes and Furui~\cite{pontes2010modeling} introduce a model of liaison from data using a decision tree learning algorithm. 


Unlike previous work, we utilize a BERT model to predict the liaison behavior of target words. We fine-tune the BERT model using liaison data labeled by a language expert, which encompasses approximately 1600 sentences and 1800 instances of liaisons. The result is a binary classification that predicts liaison occurrence. Given the intricate relationship between liaison behavior and part-of-speech (POS) in French, we concurrently train the model for the POS task using data from TreeBank\footnote{https://universaldependencies.org/}, which consists of about 14000 sentences.

\subsubsection{Polyphone}
There are numerous homophones, which is attributed to the intricate relationship between pronunciation and spelling in the French language. To achieve improved French TTS pronunciation, a polyphone model is trained based on BERT. Therefore, our final model encompasses multiple tasks, including POS tagging, polyphone disambiguation, and liaison prediction. For most polyphone words' training data, it's labeled by OpenAI text-davinci-003 model~\cite{brown2020language} with employing appropriate prompts, each polyphone word is labeled with about 1000 sentences.

\subsection{Contextual Modeling}
\label{context}
\subsubsection{Contextual Encoder}
Given the same sentence with a different context, the prosody of the generated speech would be different. Therefore, we adopt a text-based contextual encoder~\cite{xiao2023contextspeech} to model the contextual information to enhance the prosody coherence and expressiveness of long-form content reading. Given a paragraph text, for each sentence centered in a predefined number of sentences of context, our contextual encoder extracts two kinds of contextual representations: token-based and sentence-based representation. These two features are up-sampled to phoneme-level and then added to the conformer encoder input as shown in Figure \ref{fig:overview}.

Specifically, we use a BERT and statistical feature extractor to extract the token-level semantic embedding and syntactic information, which is derived as a token-based contextual representation. For the sentence-based contextual representation, we extract each sentence's BERT embedding (the average of BERT token's embedding) for the predefined range of context in the paragraph and assemble them into a matrix. The matrix is to stack sentence embeddings in the order of sentences in the paragraph. The sentence embedding matrix will be then encoded by a GRU module to generate a state vector, which contains both historical and future information. With the above design of a hierarchical contextual encoder, the Conformer encoder broadens the horizon of the current phoneme to paragraph scope, which will enhance prosody coherence and expressiveness. 

\subsubsection{Emotion Encoder}
Generally, when humans read a passage, their emotional state could change according to the context. Usually, calm is used for narration content while dialogue content would convey more emotional states such as joy, anger, and sorrow according to the context. In order to detect and express the emotion of dialogue, we adopt a context-aware emotion encoder to capture the emotional states. 

In particular, we fine-tune a pre-trained BERT with a downstream emotion classification task. To consider the contextual information, we use the context of the dialogue (add up to 256 tokens), as BERT input and use the average of token embedding of the current dialogue sentence in context as sentence embedding (CSE) to build the emotion classification model, which consists of a dense layer and cross-entropy loss. 


Our training data comes from French novel texts found on the web, which we labeled with an emotion tag using OpenAI’s text-davinci-003 model with suitable prompts. In the TTS training process, we use the CSE as an implicit emotion representation, replacing the emotion tag. This is because the implicit embedding is smoother and includes emotional intensity. After processing through a linear layer, we combine this emotion embedding with the output from the Conformer encoder, as depicted in Figure \ref{fig:overview}. For narration contents, which have no emotion, the CSE is zero-initialized.

\subsection{Conformer based Acoustic Model}
The Acoustic model we used follows the backbone in Blizzard Challenge 2021, which is a phone-level duration-based fastspeech2 liked acoustic model but changes the encoder and decoder module into the latest Conformer module. 
Besides, we add some variance information to help handle the one-to-many problem, which includes speaker ID, narration-or-dialogue ID, phone-level pitch, utterance-level Global Style Token(GST)~\cite{wang2018style}, and phone-level prosody embedding. For simplicity, we've consolidated the phone-level pitch, GST, and phone-level prosody embedding into a variance adaptor\cite{ren2020fastspeech}, as depicted in Figure\ref{fig:overview}.   

The Conformer~\cite{gulati2020conformer}, a type of Transformer, blends elements from convolutional neural networks (CNNs) and Transformer components. It effectively learns both global and local interactions through a mix of self-attention and convolution, a strategy found useful in TTS tasks.

The phone-level pitches are determined by averaging the ground truth frame-level pitch values, according to the duration of the phone sequences.

For prosody at both utterance and phoneme levels, we employ two reference encoders to extract values from reference mel during training, and two separate predictors to estimate values during inference. The utterance-level prosody vector is derived from a GST-based reference encoder, followed by a GRU and style token layer. The GST predictor at the utterance level uses a GRU and a bottleneck layer to predict the prosody vector.

Phoneme-level prosody vectors are derived from a Conformer-based reference encoder. These vectors are predicted using the phoneme encoder's outputs. The predictor for phoneme-level prosody, made up of two GRU layers, uses both text encoder outputs and the utterance-level prosody vector for hierarchical predictions.

\section{Task Description, Data Processing, and Training Strategy}

\subsection{Task Description}\label{sec:task}


Blizzard Challenge 2023 has 2 task tracks. 
\begin{itemize}
    \item FH1: 50 hours paired text-speech data with 22.05 kHz sample rate from a female native speaker of French as hub task. 
    \item FS1: 2 hours paired text-speech data with 22.05 kHz sample rate from a second female native speaker of French as spoke task. 

\end{itemize}
We build the FH1 TTS only on the given 50 hours of data and utilize a French multi-speaker source model to adapt to spoke speaker for task FS1. The multi-speaker source model for French is developed utilizing an internal corpus that comprises over 100 hours of clear French data. This corpus includes contributions from 53 speakers, with a balanced representation of genders, including 25 females and 28 males, all exhibiting a general reading speech style.

\subsection{Data Processing}

\subsubsection{Raw data segmentation}
The raw data provided by the organizer is divided into audiobook chapters. To process this data, we use the provided timestamps to divide the lengthy audio into smaller segments. To maintain the long and continuous context, we concatenate short sentences using comma and quotation punctuation, ensuring that the audio length falls within an approximate range of 5 to 20 seconds. Furthermore, concatenation would be terminated if a sentence encountered an EOS (End-Of-Sentence) mark or reached the end of a chapter.

\subsubsection{Narration/Dialogue segmentation}
Through a simple study of French dialogue format, we design several rules to detect narration and dialogue in the text as follows:
\begin{itemize}
    \item $R_1$: Within the paragraph, the content enclosed in quotation marks \guillemotleft\guillemotright \ is considered as dialogue, but if the content is too short, that might be some modal particles, etc., which are treated as the narration in our rules.
    \item $R_2$: Paragraphs starting with ¬ and ending until the current paragraph are considered as dialogue.
\end{itemize}

\subsubsection{Post-process for model training}

The French input text is converted into a phoneme sequence via sentence separation, text normalization(TN), IPA-based\footnote{https://www.internationalphoneticalphabet.org} Grapheme-to-Phoneme(G2P) model, and a French lexicon. We down-sample the audio used to 16 kHz to train the acoustic model. As the speech data contains noise, we utilize a speech enhancement tool~\cite{eskimez2023real} to reduce it, which we find advantageous for both acoustic model and vocoder training, resulting in cleaner and more natural synthesized speech. Following this, mel-spectrograms are extracted using a short-time Fourier transform (STFT) with a frame size of 50 ms, a frame hop of 12.5 ms, and a Hann window function. For vocoder training, we adhere to the strategy set out in the Blizzard Challenge 2021. We utilize the 24k audios and 16k mel-spectrograms to train the HiFi-GAN ~\cite{kong2020hifi} using all the training data mentioned in Section \ref{sec:task}, and subsequently, we separately fine-tune the model for each track. Lastly, to match the sample rate of the original recording, we down-sample the audio from 24kHz to 22.05kHz.

\subsection{Training Strategy} 
The loss function used in the MuLanTTS TTS system is shown in Equation \ref{eq:loss}
\begin{equation}
\label{eq:loss}
L = L_{gst}+L_{phone}+L_{pitch}+L_{dur}+L_{mel}+L_{ssim}
\end{equation}

The function is multifaceted and includes several key components. For instance, $L_{gst}$ represents the L1 loss between the predicted global style token for an entire utterance and the token extracted from the recording using a reference encoder. Similarly, $L_{phone}$ calculates the L1 loss between the predicted phoneme-level prosody vectors and the vectors that are extracted from the recording using a phoneme-level reference encoder. Furthermore, to calculate the L1 loss between the predicted and actual values for pitch and duration, we use $L_{pitch}$ and $L_{dur}$, respectively. The actual pitch values are obtained using the internal pitch extractor tool, while the HMM force alignment is employed to extract the actual duration values. $L_{mel}$ is the sum of mel-spectrogram L1 loss between the predicted and ground-truth mel-spectrogram in each Conformer block. We also follow the Structural Similarity measure to compare the predicted and ground-truth mel-spectrograms in the final Conformer block, resulting in $L_{ssim}$.  



The Conformer module includes 6 Conformer blocks, with an attention dimension of 512 and a hidden size of 2048 for the convolutional feed-forward module. We use the Ranger optimizer for training, with parameters set at $\alpha=6$, $k=6$, and $\epsilon=1e-5$. The initial learning rate is $1e-3$, with a warm-up period of 4000 steps and exponential decay. The training process is carried out on 8 NVIDIA V100 GPUs.

\section{Results}

The Blizzard Challenge 2023 evaluation includes four tests: a naturalness test using the Mean Opinion Score (MOS) and Multiple Stimuli with Hidden Reference and Anchor (MUSHRA); a similarity test via the Similarity Mean Opinion Score (SMOS); and two intelligibility tests, scored through the Word Error Rate (WER) and a homograph test (HOMOS). In MOS, listeners rate the naturalness of a single sample from 1 to 5. In SMOS, they rate similarity to a reference voice from 1 to 5. The intelligibility tests involve listeners typing what they hear from one-time-only audios. The audios, both synthesized and natural, are rated by three types of listeners: paid, online volunteers, and speech experts


\subsection{Naturalness Test}
\subsubsection{FH1-MOS-MUSHRA}

\begin{figure}[h]  
    \centering  
    \begin{subfigure}[h]{0.45\textwidth}  
        \centering  
        \includegraphics[width=\textwidth]{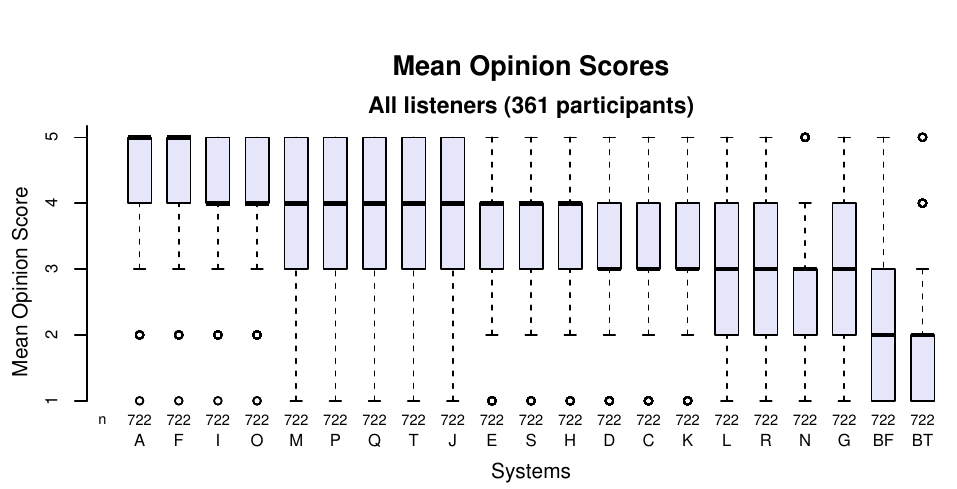}
         \caption{Boxplot of MOS in track FH1.}
        \label{fig:FH1_MOS}  
    \end{subfigure}  
    \hfill  
    \begin{subfigure}[h]{0.35\textwidth}  
        \centering  
        \includegraphics[width=\textwidth]{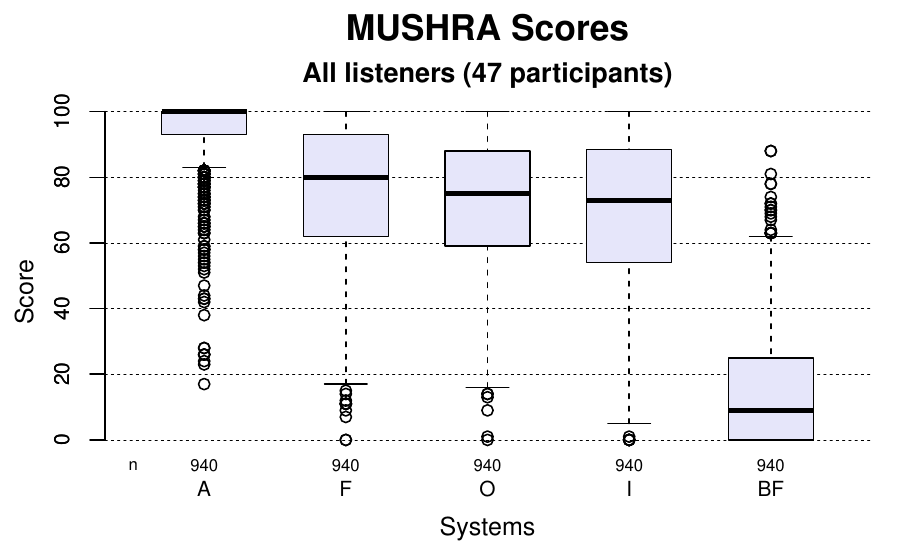}
        \caption{Boxplot of MUSHRA scores of Top-3 submitted system in track FH1.}
        \label{fig:FH1_MUSHRA}  
    \end{subfigure}  
    \caption{MOS and MUSHRA scores for naturalness test in FH1. Our system is identified as F with the recording (A) and FastSpeech baseline (BF).}  
    \label{fig:FH1_naturalness}  
\end{figure}  

Naturalness test evaluation results in FH1 task of all systems are shown in Figure \ref{fig:FH1_MOS} and \ref{fig:FH1_MUSHRA}. The MOS result of our system shows a better statistical score distribution than all the other TTS systems similar to recording, and we have a close MOS compared to recording, 4.3(F) vs 4.4(recording). Although there is a gap between the MUSHRA score of 0.75(F) and the recording score of 0.94(A), our system is also verified as the leading naturalness compared to all other synthesized systems.

\subsubsection{FS1-MOS-MUSHRA}

\begin{figure}[h!]
    \centering  
    \begin{subfigure}[h]{0.45\textwidth}  
        \centering  
        \includegraphics[width=\textwidth]{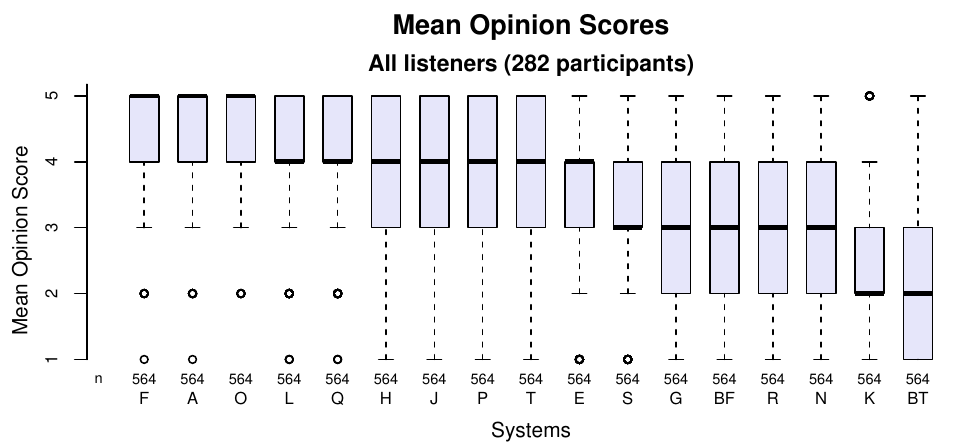}
         \caption{Boxplot of MOS in track FS1.}
        \label{fig:FS1_MOS}  
    \end{subfigure}  
    \hfill  
    \begin{subfigure}[h]{0.35\textwidth}  
        \centering  
        \includegraphics[width=\textwidth]{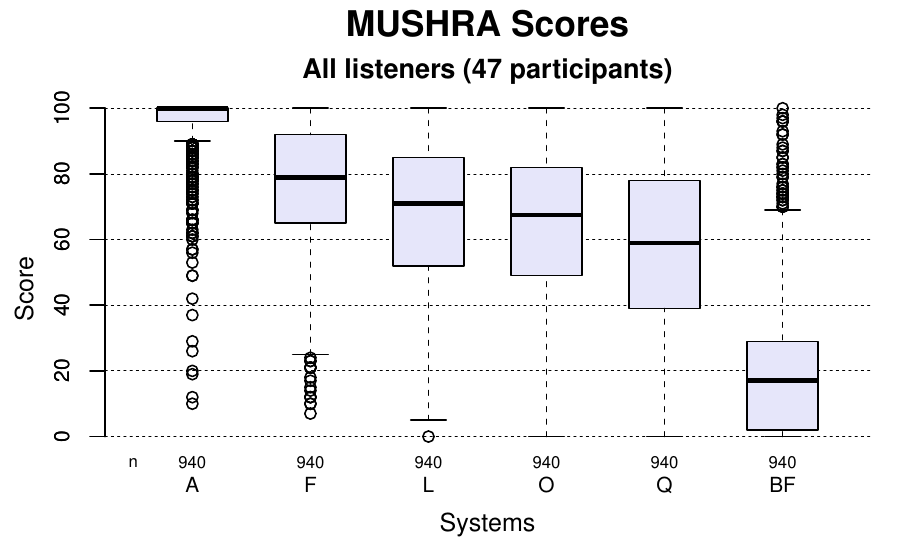}
        \caption{Boxplot of MUSHRA scores of Top-4 submitted system in track FS1.}
        \label{fig:FS1_MUSHRA}  
    \end{subfigure}  
    \caption{MOS and MUSHRA scores for naturalness test in FH1. Our system is identified as F with the recording (A) and FastSpeech baseline (BF).}  
    \label{fig:FS1_naturalness}  
\end{figure}  

The naturalness test evaluation results for the FS1 task for all systems are depicted in Figure \ref{fig:FS1_MOS} and \ref{fig:FS1_MUSHRA}. Our system achieved a higher MOS than all other TTS systems among the synthesized systems, and our MOS is comparable to that of the recording. In fact, our system is even more stable than the recording, with a score of 4.5(F) versus 4.5(recording) and std 0.70(F) versus 0.77(A). According to mean MUSHRA score 0.76(F) and the recording score of 0.95(A), it also demonstrates our system gains the highest score than other top submitted TTS systems.

\subsection{Similarity Test}

\subsubsection{FH1-SMOS}

\begin{figure}[ht]
  \centering
  \includegraphics[width=0.85\linewidth]{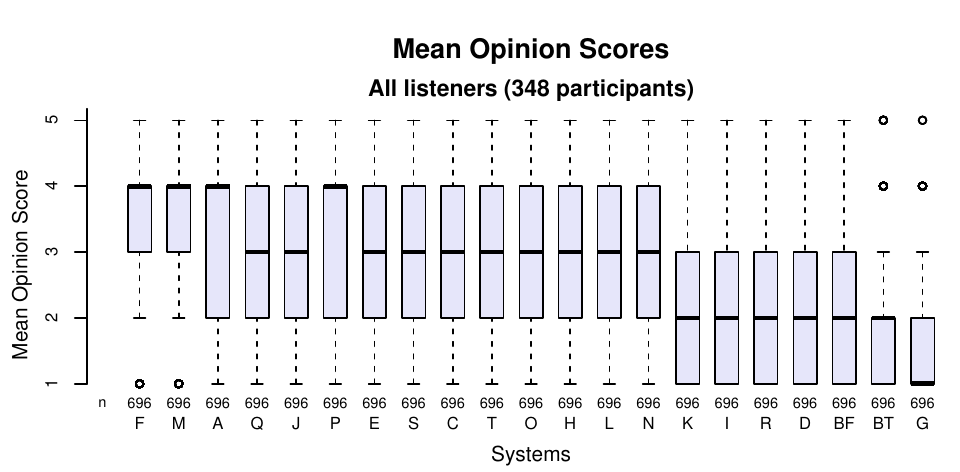}
  \caption{Boxplot of SMOS in track FH1. Our System is identified as F.}
  \label{fig:FH1_SMOS}
\end{figure}

Figure \ref{fig:FH1_SMOS} displays the MOS for similarity evaluation results of all systems, as rated by all listeners. Our System F attains the highest MOS of 3.4, which is the same as the recording. However, this is not a particularly high similarity score. We hypothesize that this could be due to the large timbre variance in the FH1 track, which may be affected by changes in acoustic environments and channels across different chapters and sections of audiobook data. The similarity test may yield low scores if the reference audio samples differ from the training data. 

\subsubsection{FS1-SMOS}

\begin{figure}[ht]
  \centering
  \includegraphics[width=0.85\linewidth]{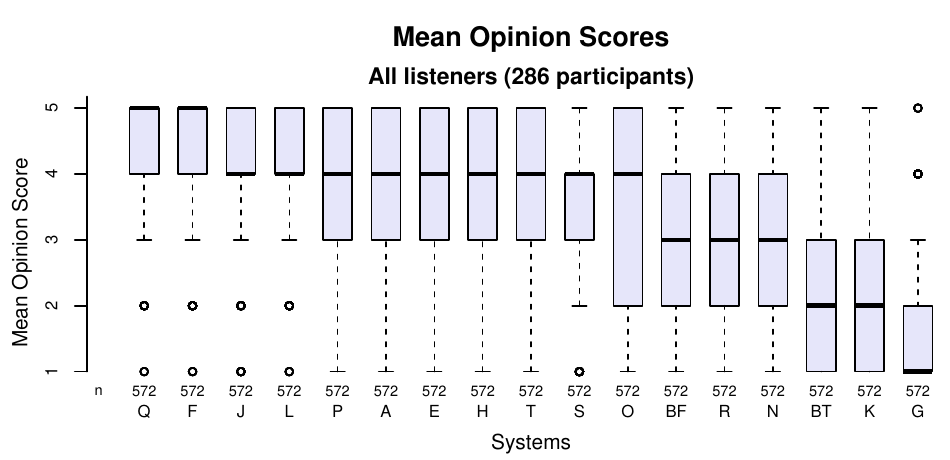}
  \caption{Boxplot of SMOS in track FS1. Our System is identified as F.}
  \label{fig:FS1_SMOS}
\end{figure}

Figure \ref{fig:FS1_SMOS} illustrates the SMOS evaluation results across all systems. Our System F attains a score of 4.3, which is statistically on par with System Q and significantly higher than recording A (4.0).

\subsection{Intelligibility of Sentences on SUS}
\begin{figure}[!]
  \centering
  \includegraphics[width=0.95\linewidth]{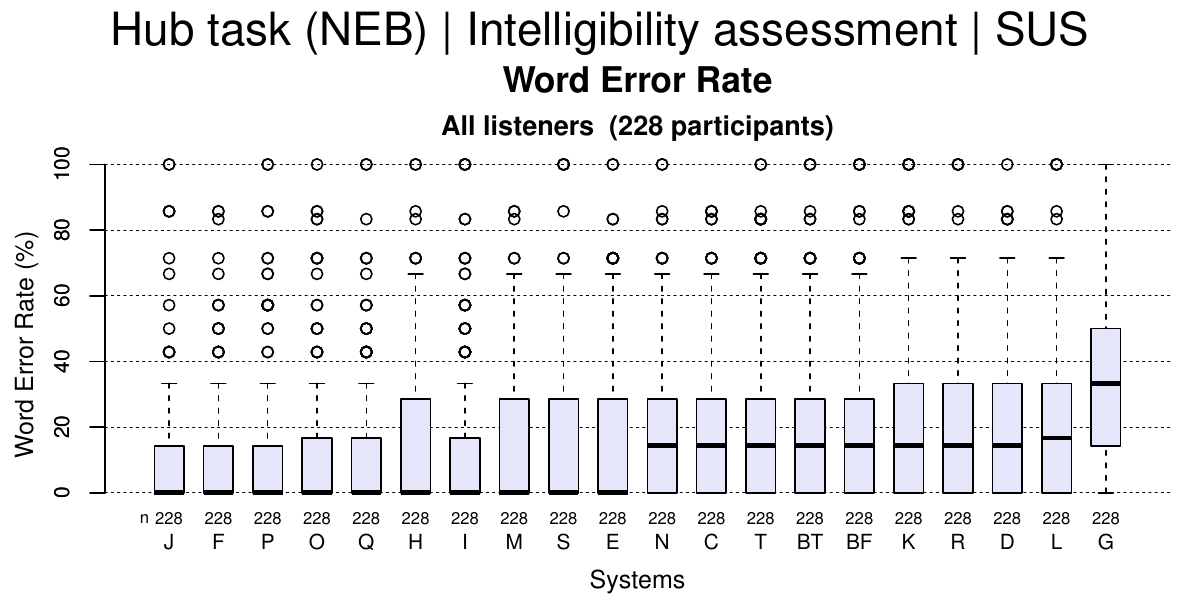}
  \caption{Word error rates for SUS test.}
  \label{fig:INT_SUS}
\end{figure}

The intelligibility test results for the SUS set, as evaluated by paid listeners and native speakers, are depicted in Figure \ref{fig:INT_SUS}. The SUS set ~\cite{benoit1996sus} was manually generated utilizing specific grammar structures for this test. In this evaluation, our proposed System F achieves the second-best mean WER (0.101), slightly trailing System J (0.095).


\subsection{Intelligibility of Sentences on Homographs}

\begin{figure}[!]
  \centering
  \includegraphics[width=0.6\linewidth]{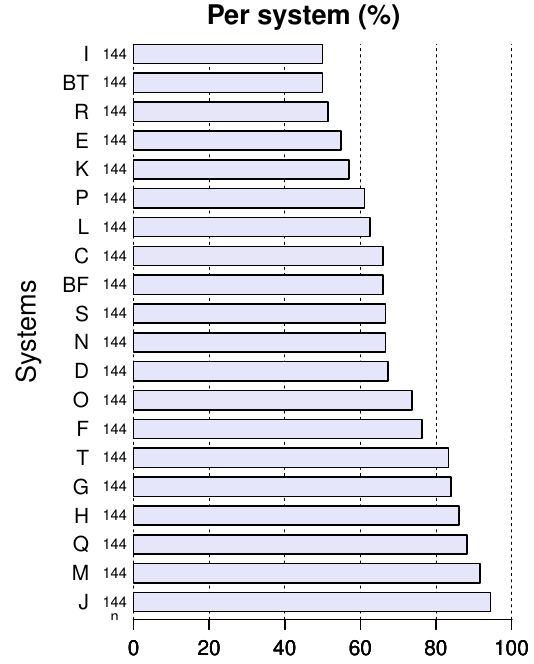}
  \caption{Accuracy for Homographs test.}
  \label{fig:int_sha}
\end{figure}

Figure \ref{fig:int_sha} shows our system F's HOMOS results, achieving nearly 80\% accuracy, which is above average. However, over 60\% of the homographs have not been addressed by our system. We plan to improve our TTS performance on these homographs for better accuracy in the future.

\section{Conclusions}


This paper introduces Microsoft's neural TTS system, MuLanTTS, developed for the Blizzard Challenge 2023 to create an audiobook speaker TTS and another spoke speaker adaptation. Built on DelightfulTTS, it uses contextual and emotion encoders to boost prosody and dialogue expressiveness. Denoising algorithms and long audio processing also enhance recording quality. MuLanTTS's performance, indicated by mean quality scores of 4.3 and 4.5, is comparable to natural speech and maintains good similarity. The superior performance of our system in this year's rigorous statistical evaluation underscores the effectiveness of the proposed system in both tasks. It is also noticed that the paragraph test samples haven't been evaluated yet in the evaluation and there are still rooms for complex content synthesis in a long context and highly expressive dialogue synthesis to improve for future work. 

\section{Acknowledgement}


The authors appreciate Huaming Wang, Kun Cong, Hanye Xu, and Min Tang for their valuable model on speech enhancement, and Gang Wang, Songze Wu, and Zheng Niu for their support and discussions on French language processing and model building. At last, we appreciate the organizers' great effort in the dataset preparation, various evaluations, and detailed statistical analysis. 

\clearpage
\bibliographystyle{IEEEtran}
\bibliography{mybib}

\begin{thebibliography}{10}
\providecommand{\url}[1]{#1}
\csname url@samestyle\endcsname
\providecommand{\newblock}{\relax}
\providecommand{\bibinfo}[2]{#2}
\providecommand{\BIBentrySTDinterwordspacing}{\spaceskip=0pt\relax}
\providecommand{\BIBentryALTinterwordstretchfactor}{4}
\providecommand{\BIBentryALTinterwordspacing}{\spaceskip=\fontdimen2\font plus
\BIBentryALTinterwordstretchfactor\fontdimen3\font minus
  \fontdimen4\font\relax}
\providecommand{\BIBforeignlanguage}[2]{{%
\expandafter\ifx\csname l@#1\endcsname\relax
\typeout{** WARNING: IEEEtran.bst: No hyphenation pattern has been}%
\typeout{** loaded for the language `#1'. Using the pattern for}%
\typeout{** the default language instead.}%
\else
\language=\csname l@#1\endcsname
\fi
#2}}
\providecommand{\BIBdecl}{\relax}
\BIBdecl

\bibitem{shen2018natural}
J.~Shen, R.~Pang, R.~J. Weiss, M.~Schuster, N.~Jaitly, Z.~Yang, Z.~Chen,
  Y.~Zhang, Y.~Wang, R.~Skerrv-Ryan \emph{et~al.}, ``Natural tts synthesis by
  conditioning wavenet on mel spectrogram predictions,'' in \emph{2018 IEEE
  International Conference on Acoustics, Speech and Signal Processing
  (ICASSP)}.\hskip 1em plus 0.5em minus 0.4em\relax IEEE, 2018, pp. 4779--4783.

\bibitem{ren2019fastspeech}
Y.~Ren, Y.~Ruan, X.~Tan, T.~Qin, S.~Zhao, Z.~Zhao, and T.-Y. Liu, ``Fastspeech:
  Fast, robust and controllable text to speech,'' \emph{Advances in neural
  information processing systems}, vol.~32, 2019.

\bibitem{ren2020fastspeech}
Y.~Ren, C.~Hu, X.~Tan, T.~Qin, S.~Zhao, Z.~Zhao, and T.-Y. Liu, ``Fastspeech 2:
  Fast and high-quality end-to-end text to speech,'' in \emph{International
  Conference on Learning Representations}, 2020.

\bibitem{black2005blizzard}
A.~W. Black and K.~Tokuda, ``The blizzard challenge-2005: Evaluating
  corpus-based speech synthesis on common datasets,'' in \emph{Ninth European
  Conference on Speech Communication and Technology}, 2005.

\bibitem{wang2017tacotron}
Y.~Wang, R.~Skerry-Ryan, D.~Stanton, Y.~Wu, R.~J. Weiss, N.~Jaitly, Z.~Yang,
  Y.~Xiao, Z.~Chen, S.~Bengio \emph{et~al.}, ``Tacotron: Towards end-to-end
  speech synthesis,'' \emph{Proc. Interspeech 2017}, pp. 4006--4010, 2017.

\bibitem{li2019neural}
N.~Li, S.~Liu, Y.~Liu, S.~Zhao, and M.~Liu, ``Neural speech synthesis with
  transformer network,'' in \emph{Proceedings of the AAAI Conference on
  Artificial Intelligence}, vol.~33, no.~01, 2019, pp. 6706--6713.

\bibitem{oord2016wavenet}
A.~van~den Oord, S.~Dieleman, H.~Zen, K.~Simonyan, O.~Vinyals, A.~Graves,
  N.~Kalchbrenner, A.~Senior, and K.~Kavukcuoglu, ``Wavenet: A generative model
  for raw audio,'' in \emph{9th ISCA Speech Synthesis Workshop}, pp. 125--125.

\bibitem{li2020robutrans}
N.~Li, Y.~Liu, Y.~Wu, S.~Liu, S.~Zhao, and M.~Liu, ``Robutrans: A robust
  transformer-based text-to-speech model,'' in \emph{Proceedings of the AAAI
  Conference on Artificial Intelligence}, vol.~34, no.~05, 2020, pp.
  8228--8235.

\bibitem{gulati2020conformer}
A.~Gulati, J.~Qin, C.-C. Chiu, N.~Parmar, Y.~Zhang, J.~Yu, W.~Han, S.~Wang,
  Z.~Zhang, Y.~Wu \emph{et~al.}, ``Conformer: Convolution-augmented transformer
  for speech recognition,'' \emph{Interspeech 2020}, 2020.

\bibitem{liu2021delightfultts}
Y.~Liu, Z.~Xu, G.~Wang, K.~Chen, B.~Li, X.~Tan, J.~Li, L.~He, and S.~Zhao,
  ``Delightfultts: The microsoft speech synthesis system for blizzard challenge
  2021,'' \emph{arXiv preprint arXiv:2110.12612}, 2021.

\bibitem{yi2022prosodyspeech}
Y.~Yi, L.~He, S.~Pan, X.~Wang, and Y.~Xiao, ``Prosodyspeech: Towards advanced
  prosody model for neural text-to-speech,'' in \emph{ICASSP 2022-2022 IEEE
  International Conference on Acoustics, Speech and Signal Processing
  (ICASSP)}.\hskip 1em plus 0.5em minus 0.4em\relax IEEE, 2022, pp. 7582--7586.

\bibitem{wang2023neural}
C.~Wang, S.~Chen, Y.~Wu, Z.~Zhang, L.~Zhou, S.~Liu, Z.~Chen, Y.~Liu, H.~Wang,
  J.~Li \emph{et~al.}, ``Neural codec language models are zero-shot text to
  speech synthesizers,'' \emph{arXiv preprint arXiv:2301.02111}, 2023.

\bibitem{betker2023better}
J.~Betker, ``Better speech synthesis through scaling,'' \emph{arXiv preprint
  arXiv:2305.07243}, 2023.

\bibitem{kim2021vits}
J.~Kim, J.~Kong, and J.~Son, ``Conditional variational autoencoder with
  adversarial learning for end-to-end text-to-speech,'' in \emph{International
  Conference on Machine Learning}.\hskip 1em plus 0.5em minus 0.4em\relax PMLR,
  2021, pp. 5530--5540.

\bibitem{tan2022naturalspeech}
X.~Tan, J.~Chen, H.~Liu, J.~Cong, C.~Zhang, Y.~Liu, X.~Wang, Y.~Leng, Y.~Yi,
  L.~He \emph{et~al.}, ``Naturalspeech: End-to-end text to speech synthesis
  with human-level quality,'' \emph{arXiv preprint arXiv:2205.04421}, 2022.

\bibitem{du2022vqtts}
C.~Du, Y.~Guo, X.~Chen, and K.~Yu, ``Vqtts: high-fidelity text-to-speech
  synthesis with self-supervised vq acoustic feature,'' in \emph{Interspeech
  2022, 23rd Annual Conference of the International Speech Communication
  Association, Incheon, Korea, 18-22 September 2022}.\hskip 1em plus 0.5em
  minus 0.4em\relax {ISCA}, 2022, pp. 1596--1600.

\bibitem{kharitonov2023speak}
E.~Kharitonov, D.~Vincent, Z.~Borsos, R.~Marinier, S.~Girgin, O.~Pietquin,
  M.~Sharifi, M.~Tagliasacchi, and N.~Zeghidour, ``Speak, read and prompt:
  High-fidelity text-to-speech with minimal supervision,'' \emph{arXiv preprint
  arXiv:2302.03540}, 2023.

\bibitem{du2023unicats}
C.~Du, Y.~Guo, F.~Shen, Z.~Liu, Z.~Liang, X.~Chen, S.~Wang, H.~Zhang, and
  K.~Yu, ``Unicats: A unified context-aware text-to-speech framework with
  contextual vq-diffusion and vocoding,'' \emph{arXiv preprint
  arXiv:2306.07547}, 2023.

\bibitem{xiao2022improving}
Y.~Xiao, X.~Wang, L.~He, and F.~K. Soong, ``Improving fastspeech tts with
  efficient self-attention and compact feed-forward network,'' in \emph{ICASSP
  2022-2022 IEEE International Conference on Acoustics, Speech and Signal
  Processing (ICASSP)}.\hskip 1em plus 0.5em minus 0.4em\relax IEEE, 2022, pp.
  7472--7476.

\bibitem{dai2019transformer}
Z.~Dai, Z.~Yang, Y.~Yang, J.~G. Carbonell, Q.~Le, and R.~Salakhutdinov,
  ``Transformer-xl: Attentive language models beyond a fixed-length context,''
  in \emph{Proceedings of the 57th Annual Meeting of the Association for
  Computational Linguistics}, 2019, pp. 2978--2988.

\bibitem{lei2023context}
S.~Lei, Y.~Zhou, L.~Chen, Z.~Wu, S.~Kang, and H.~Meng, ``Context-aware coherent
  speaking style prediction with hierarchical transformers for audiobook speech
  synthesis,'' in \emph{ICASSP 2023-2023 IEEE International Conference on
  Acoustics, Speech and Signal Processing (ICASSP)}.\hskip 1em plus 0.5em minus
  0.4em\relax IEEE, 2023, pp. 1--5.

\bibitem{xue2023m}
J.~Xue, Y.~Deng, F.~Wang, Y.~Li, Y.~Gao, J.~Tao, J.~Sun, and J.~Liang, ``M
  2-ctts: End-to-end multi-scale multi-modal conversational text-to-speech
  synthesis,'' in \emph{ICASSP 2023-2023 IEEE International Conference on
  Acoustics, Speech and Signal Processing (ICASSP)}.\hskip 1em plus 0.5em minus
  0.4em\relax IEEE, 2023, pp. 1--5.

\bibitem{xiao2023contextspeech}
Y.~Xiao, S.~Zhang, X.~Wang, X.~Tan, L.~He, S.~Zhao, F.~K. Soong, and T.~Lee,
  ``Contextspeech: Expressive and efficient text-to-speech for paragraph
  reading,'' \emph{arXiv preprint arXiv:2307.00782}, 2023.

\bibitem{devlin2018BERT}
J.~D. M.-W.~C. Kenton and L.~K. Toutanova, ``Bert: Pre-training of deep
  bidirectional transformers for language understanding,'' in \emph{Proceedings
  of NAACL-HLT}, 2019, pp. 4171--4186.

\bibitem{pan2020unified}
J.~Pan, X.~Yin, Z.~Zhang, S.~Liu, Y.~Zhang, Z.~Ma, and Y.~Wang, ``A unified
  sequence-to-sequence front-end model for mandarin text-to-speech synthesis,''
  in \emph{ICASSP 2020-2020 IEEE International Conference on Acoustics, Speech
  and Signal Processing (ICASSP)}.\hskip 1em plus 0.5em minus 0.4em\relax IEEE,
  2020, pp. 6689--6693.

\bibitem{kong2020hifi}
J.~Kong, J.~Kim, and J.~Bae, ``Hifi-gan: Generative adversarial networks for
  efficient and high fidelity speech synthesis,'' \emph{Advances in Neural
  Information Processing Systems}, vol.~33, pp. 17\,022--17\,033, 2020.

\bibitem{pontes2010modeling}
J.~d. J.~A. Pontes and S.~Furui, ``Modeling liaison in french by using decision
  trees,'' in \emph{Eleventh Annual Conference of the International Speech
  Communication Association}, 2010.

\bibitem{brown2020language}
T.~Brown, B.~Mann, N.~Ryder, M.~Subbiah, J.~D. Kaplan, P.~Dhariwal,
  A.~Neelakantan, P.~Shyam, G.~Sastry, A.~Askell \emph{et~al.}, ``Language
  models are few-shot learners,'' \emph{Advances in neural information
  processing systems}, vol.~33, pp. 1877--1901, 2020.

\bibitem{wang2018style}
Y.~Wang, D.~Stanton, Y.~Zhang, R.-S. Ryan, E.~Battenberg, J.~Shor, Y.~Xiao,
  Y.~Jia, F.~Ren, and R.~A. Saurous, ``Style tokens: Unsupervised style
  modeling, control and transfer in end-to-end speech synthesis,'' in
  \emph{International Conference on Machine Learning}.\hskip 1em plus 0.5em
  minus 0.4em\relax PMLR, 2018, pp. 5180--5189.

\bibitem{eskimez2023real}
S.~E. Eskimez, T.~Yoshioka, A.~Ju, M.~Tang, T.~Parnamaa, and H.~Wang,
  ``Real-time joint personalized speech enhancement and acoustic echo
  cancellation with e3net,'' \emph{arXiv preprint arXiv:2211.02773}, 2023.

\bibitem{benoit1996sus}
C.~Beno{\^\i}t, M.~Grice, and V.~Hazan, ``The sus test: A method for the
  assessment of text-to-speech synthesis intelligibility using semantically
  unpredictable sentences,'' \emph{Speech communication}, vol.~18, no.~4, pp.
  381--392, 1996.

\end{thebibliography}

\end{document}